# Breakdown of effective-medium theory by a photonic spin Hall effect

Shuaijie Yuan[1,†], Xinxing Zhou[1,†,*], Yu Chen[2,†], Yuhan Zhong[3,4], Lijuan Sheng[1], Hao Hu[5,6], Hongsheng Chen[3,4,7,8,*], Ido Kaminer[9], and Xiao Lin[3,4,*]

[1]*Key Laboratory of Low-Dimensional Quantum Structures and Quantum Control of Ministry of Education, Synergetic Innovation Center for Quantum Effects and Applications, School of Physics and Electronics, Hunan Normal University, Changsha 410081, China*

[2]*International Collaborative Laboratory of 2D Materials for Optoelectronics Science and Technology, Engineering Technology Research Center for 2D Material Information Function Devices and Systems of Guangdong Province, Institute of Microscale Optoelectronics, Shenzhen University, Shenzhen 518060, China*

[3]*Interdisciplinary Center for Quantum Information, State Key Laboratory of Extreme Photonics and Instrumentation, ZJU-Hangzhou Global Scientific and Technological Innovation Center, Zhejiang University, Hangzhou 310027, China*

[4]*International Joint Innovation Center, The Electromagnetics Academy at Zhejiang University, Zhejiang University, Haining 314400, China*

[5]*School of Electrical and Electronic Engineering, Nanyang Technological University, 50 Nanyang Avenue, Singapore 639798, Singapore*

[6]*Key Laboratory of Radar Imaging and Microwave Photonics, Ministry of Education, College of Electronic and Information Engineering, Nanjing University of Aeronautics and Astronautics, Nanjing 211106, China*

[7]*Key Lab. of Advanced Micro/Nano Electronic Devices & Smart Systems of Zhejiang, Jinhua Institute of Zhejiang University, Zhejiang University, Jinhua 321099, China*

[8]*Shaoxing Institute of Zhejiang University, Zhejiang University, Shaoxing 312000, China*

[9]*Department of Electrical Engineering, Technion-Israel Institute of Technology, Haifa 32000, Israel*

[†]*These authors contributed equally to this work*

*Corresponding authors: xinxingzhou@hunnu.edu.cn (X. Zhou); hansomchen@zju.edu.cn (H. Chen); xiaolinzju@zju.edu.cn (X. Lin)

**Effective-medium theory pertains to the theoretical modelling of homogenization, which aims to replace an inhomogeneous structure of subwavelength-scale constituents with a homogeneous effective medium. The effective-medium theory is fundamental to various realms, including electromagnetics and material science, since it can largely decrease the complexity in the exploration of light-matter interactions by providing simple acceptable approximation. Generally, the effective-medium theory is thought to be applicable to any all-dielectric system with deep-subwavelength constituents, under the condition that the effective medium does not have a critical angle, at which the total internal reflection occurs. Here we reveal a fundamental breakdown of the effective-medium theory that can be applied in very general conditions: showing it for deep-subwavelength all-dielectric multilayers even without critical angle. Our finding relies on an exotic photonic spin Hall effect, which is shown to be ultra-sensitive to the stacking order of deep-subwavelength dielectric layers, since the spin-orbit interaction of light is dependent on slight phase accumulations during the wave propagation. Our results indicate that**

**the photonic spin Hall effect could provide a promising and powerful tool for measuring structural defects for all-dielectric systems even in the extreme nanometer scale.**

1 Introduction

Light-matter interaction lies at the heart of modern photonics and electromagnetics, and it can be rather complex, especially in inhomogeneous structures or random media. Albeit inhomogeneous, when the constituent materials have their dimension in the subwavelength or deep-subwavelength scales, the corresponding structures can be readily homogenized by exploiting the effective-medium theory [1-4] (Fig. 1a). The effective-medium theory is thus a fundamental approximation method for finding the simple homogeneous material that provides the exact same response as the complex inhomogeneous structure. At that atomic scale, every material is mostly vacuum, with only a tiny fraction of space filled with atoms. And yet, most materials can be described by a macroscopic property such as index of refraction, permittivity, permeability, or surface conductivity [5-8]. This well-known result is itself the result of the effective-medium theory. Thus, this theory lies at the foundations of optics, photonics, and electromagnetism.

The development of the effective-medium theory can be traced back to the earliest days of electromagnetic wave theory and to the Maxwell-Garnet effective medium model [4,9]. Since the theory can largely decrease the complexity in optical analysis, it has been widely adopted in the exploration of light-matter interactions and has led to many significant findings, including the material design as varied as metamaterials [10-13], metasurfaces [14-17], negative-index materials [18-21], to hyperbolic materials [22-24], and the applications in cloak [25,26] and super-lens [27-30].

Despite its ubiquitous applications, the effective-medium theory fails to describe the homogenization of metal-based structures involving modes with their wave-vector much larger than the wave-vector in free space, such as metal plasmons and hyperbolic high-$k$ modes in periodic metal-dielectric structures [31-34]. Several years ago, it was found that counterintuitively, the effective-medium theory could also break in all-dielectric structures [35-39]. This result was surprising at the time because such materials generally do not involve modes with extremely large wave-vectors. Instead, the effective-medium theory breaks near a critical angle of incidence, at which the total internal reflection occurs. For such an angle to exist for light incident from outside the structure, the superstrate must have a relative permittivity larger than that of the structure, and thus usually larger the unity. Thus, the environment cannot be free space. This requirement makes the breakdown of the effective-medium theory much harder to achieve in practice. Consequently, the effective-medium theory is generally believed to be applicable in all cases in which the all-dielectric structure is in free space, or for any all-dielectric structure covered by a superstrate with an index lower than that of the constituents of the structure. More generally, it was thought that any deep-subwavelength-scale structure that is all-dielectric can be homogenized when it has no critical angles.

Here, we find an exotic photonic spin Hall effect that fundamentally breaks the effective-medium theory. This breakdown can occur in free space, going beyond the consensus about the effective-medium theory and the conditions for its breakdown (applicable for all-dielectric structures without a critical angle). The photonic spin Hall effect [40-47] originates from the spin-orbit interaction of light and relates to a spin-dependent splitting phenomenon of the left-/right-handed circularly polarized components of a reflected light beam. We show that the photonic spin Hall effect is ultrasensitive to the realistic structural details of all-dielectric multilayers, such as the stacking order of deep-subwavelength dielectric layers, even when the thickness of each layer is down to one nanometer (Figs. S6-S7). The underlying mechanism is that the spin-orbit interaction of light [48-56] is determined by the phase accumulation during wave propagation, in addition to the amplitude of reflected light. Through this phase accumulation, our work reveals a breakdown of effective-medium theory for the homogenization of all-dielectric structures far away from the critical angle. We further find that the breakdown can also occur in ultrathin all-dielectric structures of just a few layers, going down to just a pair of layers with a total thickness in the nanometer scale (see Appendix). Due to the ultrathin thickness of each constituent layers investigated in this work, our finding indicates the possible breakdown of effective-medium theory even for atomic-scale layers, by exploiting the photonic spin Hall effect. This simplicity of our findings and their fundamental nature will facilitate a direct short-term experimental verification.

## 2 Results and discussion

For illustration, a one-dimensional periodic multilayer composed of two transparent dielectrics and with a periodicity number of $N_{pair} \gg 1$ is shown in Fig. 1a. One constituent dielectric has a relatively-high relative permittivity of $\varepsilon_H = 9$ and a thickness of $d_H = 5$ nm, and the other has a relatively-low relative permittivity of $\varepsilon_L = 4$ and a thickness of $d_L = 5$ nm. According to the realistic stacking orders, the multilayer in Fig. 1a is denoted either as $[HL]^{N_{pair}}$ or $[LH]^{N_{pair}}$. Since $d_L = d_H \ll \lambda$, the deep-subwavelength all-dielectric multilayers of both $[HL]^{N_{pair}}$ and $[LH]^{N_{pair}}$ can be modelled by a same effective uniaxial medium with a relative permittivity of $\bar{\bar{\varepsilon}}_{EMT} = [\varepsilon_\parallel, \varepsilon_\parallel, \varepsilon_\perp]$, where $\lambda = 500$ nm is the wavelength of light in free space. According to the effective-medium theory, $\bar{\bar{\varepsilon}}_{EMT}$ is irrelevant to the optical response of the surrounding media, and we have

$$\varepsilon_\parallel = \frac{\varepsilon_H d_H + \varepsilon_L d_L}{d_H + d_L} = 6.50 \quad \text{and} \quad \varepsilon_\perp = \frac{\varepsilon_L \varepsilon_H (d_H + d_L)}{\varepsilon_L d_H + \varepsilon_H d_L} = 5.54.$$

To check the validity of effective-medium theory, we consider the photonic spin Hall effect under the incidence of Gaussian beam (see the schematic in the left panel of Fig. 1b), while Ref. [35] considers the reflectance under the incidence of plane wave (right panel of Fig. 1b). Here we set the incident light to be *p*-polarized. To facilitate the comparison between this work and Ref. [35], the structural setup in Fig. 1b is similar to that in Ref. [35]. To be specific, the all-dielectric multilayer in Fig. 1b is supported by a dielectric substrate with a relative permittivity of $\varepsilon_{sub} = \varepsilon_\perp$ and covered by a dielectric superstrate with a relative permittivity of $\varepsilon_{super} = 9 > \varepsilon_\perp$. Under this scenario, the critical angle $\theta_c = 51.6°$ exists at the interface between the superstrate and the effective medium.

We now start to quantitatively investigate the photonic spin Hall effect from the all-dielectric multilayer. The photonic spin Hall effect is featured with a transverse spin-shift, which is a manifestation of the splitting of the left- and right-handed circularly polarized components of the reflected Gaussian beam in the transverse direction. According to the three-dimensional rotation matrix, the transverse spin-shift $\delta_y$ for reflected light can be approximately written as [57-59]:

$$\delta_y = -\frac{\lambda}{2\pi} \left[ 1 + \frac{|r_s|}{|r_p|} \cos(\varphi_s - \varphi_p) \right] \cot \theta_i \tag{1}$$

where $\theta_i$ is the incident angle, and

$$r_p = |r_p| e^{i\varphi_p} \tag{2}$$

$$r_s = |r_s| e^{i\varphi_s} \qquad (3)$$

That is, $|r_p|$ ($|r_s|$) and $\varphi_p$ ($\varphi_s$) denote the amplitude and phase of reflection coefficients under the incidence of *p*-polarized (*s*-polarized) plane waves from the all-dielectric multilayer, respectively; see derivation in supporting section S1.

Based on equation (1), Fig. 1c shows the transverse spin-shift as a function of the incident angle from three structures, namely $\delta_y^{HL}$ from the realistic multilayer of $[HL]^{N_{pair}}$, $\delta_y^{LH}$ from the realistic multilayer of $[LH]^{N_{pair}}$, and $\delta_y^{EMT}$ from the homogenized effective medium. Large discrepancies are found between these transverse spin-shifts in a wide range of incident angles in Fig. 1c. For example, the maximum value of $|\Delta_\delta| = |\delta_y^{LH} - \delta_y^{HL}|$ is close to 1 μm. Due to these large discrepancies, we can conclude that the photonic spin Hall effect is sensitive to the realistic stacking orders of all-dielectric multilayers. In other words, the effective-medium theory is no longer precise to model the homogenization process of these all-dielectric multilayers, despite the deep-subwavelength size of their constituents.

Counterintuitively, the breakdown of effective-medium theory under the incidence of Gaussian beam appears not only near the critical angle (namely $\theta_i \in [51.6°, 51.7°]$) but also far away from the critical angle (namely $\theta_i \in [18°, 28°]$ or $\theta_i \in [34°, 41°]$), as highlighted by three grey regions in Fig. 1c. For example, the maximum value of $|\Delta_\delta|$ in Fig. 1c appears at $\theta_{i,max} = 22.5°$, which is far smaller than the critical angle (namely $\theta_{i,max} \ll \theta_c$). By contrast, the breakdown of effective-medium theory under the incidence of plane wave appears only close to the critical angle (namely $\theta_i \in [50.6°, 51.6°]$) in Fig. 1d, as firstly reported in Ref. [35]. In other words, according to all previous works [35-39], the effective-medium theory still works well when the incident angle is far away from the critical angle. Therefore, the breakdown of effective-medium theory away from the critical angle reported in this work has never been discovered before.

The breakdown of effective-medium theory beyond the critical angle is closely related to the fact that the photonic spin Hall effect is dependent not only on the magnitude of reflected light but also on the phase accumulation during wave propagation. To be specific, the transverse spin-shift is not only a function of the magnitudes of reflection coefficients (namely $|r_p|$ and $|r_s|$) but also a function of their phases ($\varphi_p$ and $\varphi_s$), as clearly indicated in equation (1). Regarding to the magnitudes of reflection coefficients, the breakdown of effective-medium theory happens only close to the critical angle, as shown in Fig. 1d. By contrast, regarding to the phases of reflection coefficients, the breakdown of effective-medium theory can happen far away from the critical angle, as shown in Fig. S2. We emphasize that the phase accumulation during light propagation can occur in two ways, either in a continuous way caused by the propagation of light inside a bulk medium or in an abrupt way induced by the reflection/transmission of light across an interface. The total phase accumulation caused by the abrupt way in principle can be sensitive to the realistic stacking orders of all-dielectric multilayers for arbitrary incident angles, while the total phase accumulation caused by the continuous way is indeed insensitive to the stacking orders. As a result, in terms of photonic spin Hall effect which simultaneously involves the magnitude and phase of reflected light, the breakdown of effective-medium theory is in principle possible for arbitrary incident angles, as exemplified by the results in Fig. 1c.

According to previous works [35-39], one mandatory condition for the breakdown of effective-medium theory for all-dielectric systems is the existence of critical angle, which further requires the superstrate to have a relative permittivity of $\varepsilon_{super} > \varepsilon_\perp$. Since $\varepsilon_\perp > 1$ for all-dielectric multilayers, the superstrate cannot be free space but is restricted to some dielectric solids or liquids. This mandatory condition would unavoidably cause some limitation in terms of practical applications such as disorder detection [38,39]. Therefore, it is highly wanted to get rid of this mandatory condition. Correspondingly, one fundamental question is whether the breakdown of effective-medium theory can happen in all-dielectric multilayers without the existence of critical angle.

To tackle this issue, we further investigate the photonic spin Hall effect by covering the all-dielectric multilayers simply with free space, namely $\varepsilon_{super} = 1 < \varepsilon_\perp$, as schematically shown in Fig. 2a. Figure 2b shows the transverse spin-shift is ultrasensitive to the realistic stacking orders of all-dielectric multilayers, especially when the incident angle is close to $\theta_{i,|r_p|\to 0} = 69.5°$. When the light is incident with an angle around $\theta_{i,|r_p|\to 0}$, we have $|r_p| \to 0$ and therefore, the value of $\frac{|r_s|}{|r_p|}$ in equation (1) can be extremely large. Correspondingly, the transverse spin-shift near $\theta_{i,|r_p|\to 0}$ becomes even more sensitive on the phase of reflected light. As a result, Fig. 2b shows that the maximum discrepancy of $|\delta_y^{LH}-\delta_y^{HL}|$ reaches 0.6 μm near $\theta_{i,|r_p|\to 0}$, while the maximum value of $|\delta_y^{LH}|$, $|\delta_y^{HL}|$, and $|\delta_y^{EMT}|$ is only 0.55 μm. In other words, the relative deviation has $\frac{\max(|\delta_y^{LH}-\delta_y^{HL}|)}{\max(|\delta_y^{LH}|,|\delta_y^{HL}|,|\delta_y^{EMT}|)} > 100\%$, which is very large and greatly facilitates the measurement in future experiments. The large deviation in Fig. 2b indicates the severe breakdown of effective-medium theory for all-dielectric multilayers near the specific angle of $\theta_{i,|r_p|\to 0}$, without the existence of critical angle.

For comparison, Fig. 2c shows the magnitude of reflectivity. As expected, the deviation between reflectivity from $[LH]^{N_{pair}}$ and $[HL]^{N_{pair}}$ is always small (namely $|\Delta_r|=||r_p^{LH}|^2-|r_p^{HL}|^2|<1.5\%$ in Fig. 2c). The results in Fig. 2c confirm that regarding to the reflectivity, the effective-medium theory works well with acceptable approximation for arbitrary incident angles, if there is no critical angle. Based on Fig. 2b&c, we can safely argue that the results in Fig. 2 provide the first evidence showing the breakdown of effective-medium theory for all-dielectric multilayers without the existence of critical angle by exploiting the photonic spin Hall effect, completely contrary to previous cognition.

According to Ref. [35], the performance of effective-medium theory is also related to the periodicity number $N_{pair}$ of all-dielectric multilayers. That is, the severe breakdown of effective-medium theory generally happens only when $N_{pair} \gg 1$; see more discussion about the influence of $N_{pair}$ on the validity of effective-medium theory in Fig. S5. The requirement of $N_{pair} \gg 1$ would increase the complexity in practical nano-fabrications. To mitigate the issue, it is wanted to reduce $N_{pair}$, for example, down to one. We show in Fig. 3 that the breakdown of effective-medium theory can happen even if $N_{pair}=1$ by exploiting the photonic spin Hall effect. Moreover, upon close inspection, we find that the substrate permittivity can be optimized to further maximize the discrepancy of $|\delta_y^{LH}-\delta_y^{HL}|$, and the corresponding optimal permittivity is $\varepsilon_{sub} = \tan^2 \theta_{i,|r_p|\to 0}$; see analysis in Fig. S4. For illustration, we re-set $\theta_{i,|r_p|\to 0} = 68.9°$ in Figs. 3-4 and accordingly, $\varepsilon_{sub} = 6.7$. Under this scenario, we have $\max(|\delta_y^{LH} - \delta_y^{HL}|) = 3.2$ μm with $N_{pair} = 1$ near the specific angle of $\theta_{i,|r_p|\to 0}$ in Fig. 3, which is even larger than that of 0.6 μm with $N_{pair} \gg 1$ in Fig. 2.

To further facilitate the understanding of Fig. 3, we show in Fig. 4a-b the intensity distribution of the left-handed circularly-polarized component of the reflected Gaussian beam when $\theta_i = \theta_{i,|r_p|\to 0}$. The centroid position of the reflected beam from $[LH]^1$ in Fig. 4a is opposite to that from $[HL]^1$ in Fig. 4b. To be specific, for the transverse spin-shift in the y direction, we have $\delta_y^{LH} = +1.59$ μm in Fig. 4a but $\delta_y^{HL} = -1.58$ μm in Fig. 4b. Moreover, Fig. 4a&b also shows the emergence of nonzero in-plane spin-shift in the x direction. That is, we have $\delta_x^{LH} = -1.93$ μm in Fig. 4a and $\delta_x^{HL} = +2.23$ μm in Fig. 4b.

The emergence of in-plane spin-shift indicates another feasible route to demonstrate the performance of effective-medium theory for the homogenization of deep-subwavelength all-dielectric multilayers. Accordingly, we show in Figs. 4c-d these in-plane spin-shifts from the realistic multilayers of $[LH]^1$ or $[HL]^1$ and the homogenized effective medium. Remarkably, the discrepancy of $|\delta_x^{LH} - \delta_x^{HL}|$ can be very large, and we have $\max(|\delta_x^{LH} - \delta_x^{HL}|) = 5.0$ μm with $N_{pair} = 1$ near the specific angle of $\theta_{i,|r_p|\to 0}$ in Figs. 4c-d. These opposite drifts of optical centroid and large spin-shifts in Figs. 3-4 are beneficial for the experimental observation of the breakdown of effective-medium theory far away from the critical angle, without resorting to the usage of weak measurements [42,60] or other amplification mechanisms.

Considering that there will be some inevitable losses in the preparation of multilayer films, the dielectric constants are often not in the form of pure real numbers and the imaginary part appears. Generally, the loss of prepared materials is very tiny, so we focus on the range of imaginary part $\text{im}(\varepsilon_{L/H})$ from 0.000 to 0.005. Figure 5a shows the influence of imaginary part of dielectric constant on the effective-medium theory. It is found that, when the optical system is located near $\theta_{i,|r_p|\to 0} = 68.9°$, there also exists a very large shift deviation ($\max(|\delta_y^{LH} - \delta_y^{HL}|) \approx 3.3$ μm). Then, we choose the imaginary part of the dielectric constant as $\text{im}(\varepsilon_{L/H}) = 0.001$ and investigate the transverse spin-shift changing with the incident angle. Here, one realistic dielectric has a relatively-low relative permittivity of $\varepsilon_L = 4 + 0.001\text{i}$, the other has a relatively-high relative permittivity of $\varepsilon_H = 9 + 0.001$, and the anisotropic effective medium is still calculated by effective-medium theory as $\varepsilon_\parallel = \frac{\varepsilon_H d_H + \varepsilon_L d_L}{d_H + d_L} = 6.50 + 0.001\text{i}$ and $\varepsilon_\perp = \frac{\varepsilon_L \varepsilon_H (d_H + d_L)}{\varepsilon_L d_H + \varepsilon_H d_L} \approx 5.54 + 0.001\text{i}$. We can clearly find that the photonic spin Hall effect is still very sensitive to the actual stacking order of the multilayer film. As such, the possible optical loss caused by the actual preparation of the sample has not impacted the breakdown of effective-medium theory.

For practical experiment, we can introduce Stokes parameter measurement [16] or weak measurement technique [42,60] for verifying the idea of this work. Firstly, we make the corresponding samples with realistic stacking orders $[HL]^{N_{pair}}$ or $[LH]^{N_{pair}}$. Then, we consider the polarized Gaussian beam being incident at the interface of the sample and detect the transverse spin-shifts of the reflected light beam. If the spin-shift is relatively large, we can use a quarter-wave plate and a polarizer to analyze the Stokes parameter $S_3$ of reflected light beam. Here, the information of left- and right-handed circularly polarized components can be directly obtained. For measuring very tiny spin-shifts, we can introduce the weak measurement method. Here, by choosing almost perpendicular state of preselection and postselection, the tiny spin-shifts can be enhanced indirectly with appropriate amplified factor. Finally, the effective-medium theory breakdown can be observed by comparing the spin-shifts on actual structures with opposite stacking orders.

## 3 Conclusion

In summary, we have revisited the validity of effective-medium theory for deep-subwavelength all-dielectric multilayers, and revealed the possible breakdown of effective-medium theory beyond the critical angle by exploiting the photonic spin Hall effect. The underlying mechanism is that the photonic spin Hall effect can be ultrasensitive to the phase accumulation during light propagation, in addition to the magnitude of reflected light. Correspondingly, the breakdown of effective-medium theory in this work no longer relies on the surrounding environment (namely the superstrate can now also be free space) and the periodicity number of all-dielectric multilayer (namely the periodicity number can reduce to one). Moreover, the breakdown of effective-medium theory from a photonic spin Hall effect could occur even when the thickness of each dielectric layer is down to the nanometer scale and thus is also possible for atomic-scale layers. Therefore, our finding provides a brand-new insight for the effective-medium theory and shows that the photonic spin Hall effect holds great potential in the precise metrology, such as the detection of defects in deep-subwavelength all-dielectric structures in the extreme nanoscale or atomic-scale.

**Acknowledgments**
X.L. acknowledges the support partly from the National Natural Science Fund for Excellent Young Scientists Fund Program (Overseas) of China, the National Natural Science Foundation of China (NSFC) under Grant No. 62175212, Zhejiang Provincial Natural Science Fund Key Project under Grant No. Z23F050009, the Fundamental Research Funds for the Central Universities (2021FZZX001-19), and Zhejiang University Global Partnership Fund. H.C acknowledges the support partly from the Key Research and Development Program of the Ministry of Science and Technology under Grants No. 2022YFA1404704, 2022YFA1404902, and 2022YFA1405200, and the National Natural Science Foundation of China (NNSFC) under Grants No.11961141010 and No. 61975176. I.K. acknowledges the support from the Israel Science Foundation under Grant No. 3334/19 and the Israel Science Foundation under Grant No. 830/19. X.Z. was supported by the National Natural Science Foundation of China (11604095) and the Training Program for Excellent Young Innovators of Changsha (kq2107013).

**Author contributions**
All authors contributed extensively to this work. X.Z. and X.L. initiated the idea. S.Y., Y.C., and Y.Z. performed the calculation. L.S., H.C., H.H., and I.K. analyzed data and interpreted detailed results. S.Y., X.Z., I.K., and X.L. wrote the manuscript with input from the others. X.Z., H.C., I.K., and X.L. supervised the project.

**Competing interests**
The authors declare no competing interests.

**Appendix**
**Reflection coefficients.** The reflection coefficients from all-dielectric multilayers are rigorously obtained based on the transfer-matrix theory; see the analytical derivation in supplementary section S1 and Fig. S1. To consider the total phase accumulation during wave propagation, the reflection phases $\varphi_p$ and $\varphi_s$ from realistic multilayers are calculated and compared. If the superstrate has $\varepsilon_{\text{super}} = 9 > \varepsilon_\perp$, the detailed analysis is provided in supplementary sections S2 and Fig. S2. If the superstrate has $\varepsilon_{\text{super}} = 1 < \varepsilon_\perp$, the related discussion is shown supplementary sections S2 and Fig. S3.

**Influence of $\varepsilon_{sub}$ on the performance of effective-medium theory.** The relative permittivity of substrate with $\varepsilon_{sub} = \varepsilon_\perp$ is chosen in Figs. 1-2. Without loss of generality, the influence of $\varepsilon_{sub}$ on the performance of effective-medium theory is systematically analyzed in supplementary section S3 and Fig. S4. Figure S4 shows that the value of $|\delta_y^{LH} - \delta_y^{HL}|$ can be maximized through the optimization of $\varepsilon_{sub}$, and the optimal value is $\varepsilon_{sub} = \tan^2 \theta_{i,|r_p|\to 0} = 6.7$. According to this result, $\varepsilon_{sub} = \tan^2 \theta_{i,|r_p|\to 0}$ is re-set in Figs. 3-4.

**Influence of $N_{pair}$ on the performance of effective-medium theory.** For illustration, the periodicity number of the all-dielectric multilayer is set to be $N_{pair} = 50 \gg 1$ in Figs. 1-2. Without loss of generality, the influence of $N_{pair}$ on the performance of effective-medium theory is systematically investigated in supplementary section S4 and Fig. S5. Figure S5 shows that the value of $|\delta_y^{LH} - \delta_y^{HL}|$ would oscillate periodically as $N_{pair}$ increases, and the breakdown of effective-medium theory in principle can happen for arbitrary value of $N_{pair}$, even if $N_{pair}$ reduces to one as shown in Figs. 3-4.

**Influence of the thickness $d_{unit} = d_H + d_L$ of unit cell on the performance of effective-medium theory.** For conceptual demonstration, $d_H / \lambda = d_L / \lambda = 0.01$ and $d_{unit} / \lambda = 0.02$ are set in Figs. 1-4, where $\lambda = 500$ nm. We further investigate the influence of $d_{unit}$ on the performance of effective-medium theory in supplementary section S5 and Figs. S6-S7. We find the possible breakdown of effective-medium theory even if $d_{unit} / \lambda$ reduces to 0.001. That is, the breakdown of effective-medium theory from the photonic spin Hall effect could occur even when the thickness of each dielectric layer is less than one nanometer. As such, the photonic spin Hall effect can be a potential technique for detecting the dielectric structure defects on the extreme nanoscale, see Fig. S10.

**More discussion on equation (1).** The approximate expression of equation (1) should be modified if we have $|r_p| \to 0$ by considering more high-order terms of $r_p$ in the Taylor expansion; see detailed analysis in supplementary section S1.

**Photonic spin Hall effect under the incidence of s-polarized Gaussian beam.** For conceptual demonstration, the incident Gaussian beam is *p*-polarized in Figs. 1-4. For comparison, we also consider the incidence of *s*-polarized Gaussian beam in supplementary section S6. If the superstrate has $\varepsilon_{super} = 9 > \varepsilon_\parallel$, the breakdown of effective-medium theory beyond the critical angle can be clearly seen in Fig. S8. By contrast, if the superstrate has $\varepsilon_{super} = 1$, the breakdown of effective-medium theory beyond the critical angle becomes not that obvious in Fig. S9.

**Data availability**

All relevant data are available from the corresponding authors upon reasonable request.

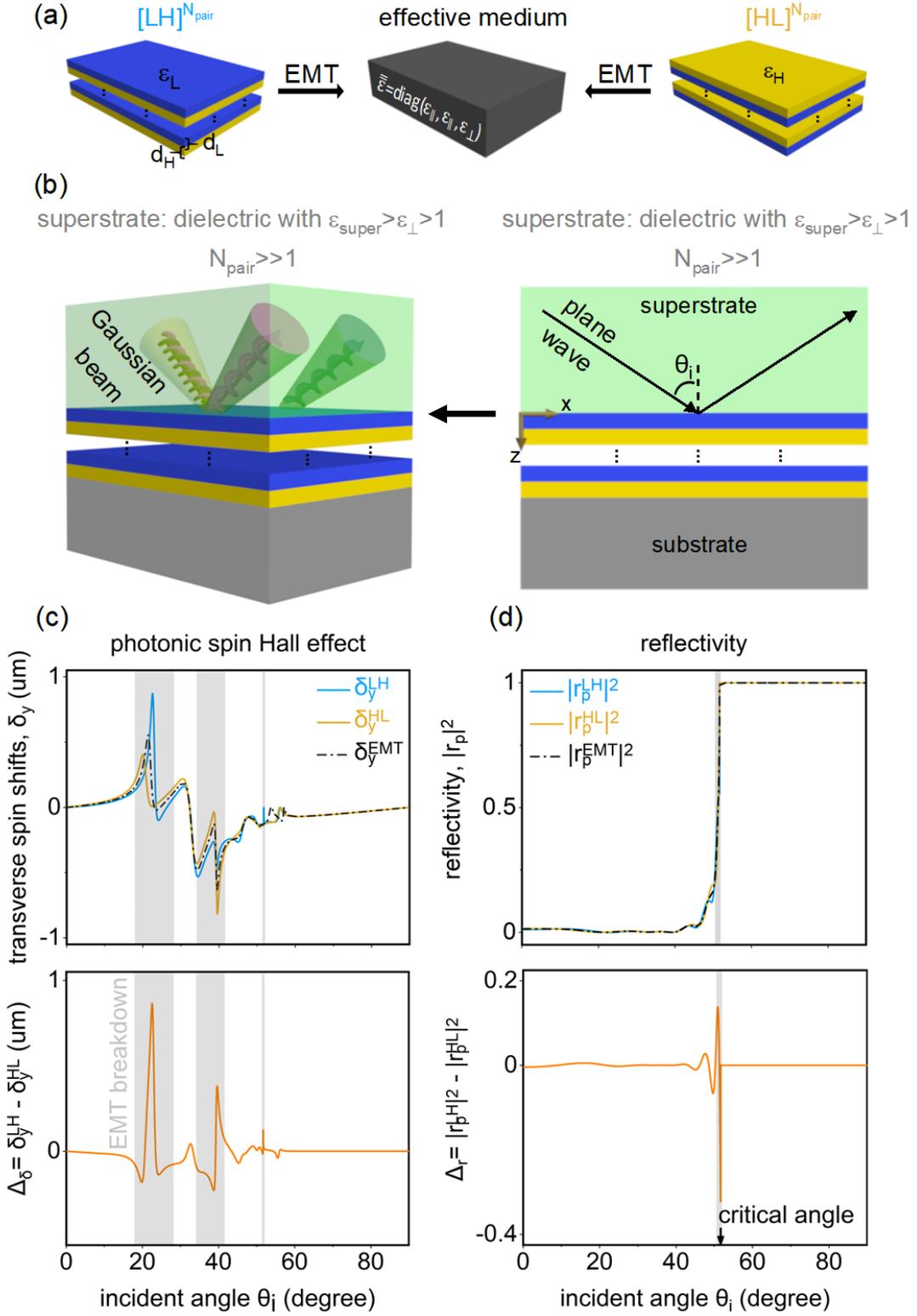

**FIG. 1. Breakdown of effective-medium theory from a photonic spin Hall effect far away from the critical angle.** Here and below, we consider the incidence of *p*-polarized wave, and the working wavelength in free space is $\lambda = 500$ nm. (**a**) Schematic of the periodic deep-subwavelength all-dielectric multilayer, composed of two transparent dielectrics. One constituent dielectric has a

relatively-low relative permittivity of $\varepsilon_L = 4$, and the other has a relatively-high relative permittivity of $\varepsilon_H = 9$. The multilayer has a periodicity number of $N_{pair} \gg 1$, such as $N_{pair} = 50$ used in Figs. 1-2. Since each constituent layer has a thickness of $d_L/\lambda = d_H/\lambda = 0.01$, the deep-subwavelength all-dielectric multilayers with different stacking orderings (e.g. $[LH]^{N_{pair}}$ in the left panel and $[HL]^{N_{pair}}$ in the right panel) are equivalent to a homogenized effective medium with a relative permittivity of $[\varepsilon_\parallel, \varepsilon_\parallel, \varepsilon_\perp]$, where $\varepsilon_\parallel = 6.50$ and $\varepsilon_\perp = 5.54$, according to the effective-medium theory. (**b**) Schematic of the photonic spin Hall effect and reflectivity from the all-dielectric multilayer, whose substrate has a relative permittivity of $\varepsilon_{sub} = \varepsilon_\perp$ and whose superstrate has a relative permittivity of $\varepsilon_{super} > \varepsilon_\perp$. (**c**) Transverse shift $\delta_y$ of photonic spin Hall effect from the realistic multilayers and their effective media under the incidence of a Gaussian beam. (**d**) Reflectivity from the realistic multilayers and their effective media under the incidence of plane waves. The grey regions in (c, d) highlight the angular ranges, within which the photonic spin Hall effect or the magnitude of reflectivity is sensitive to the stacking orders of the deep-subwavelength all-dielectric multilayer.

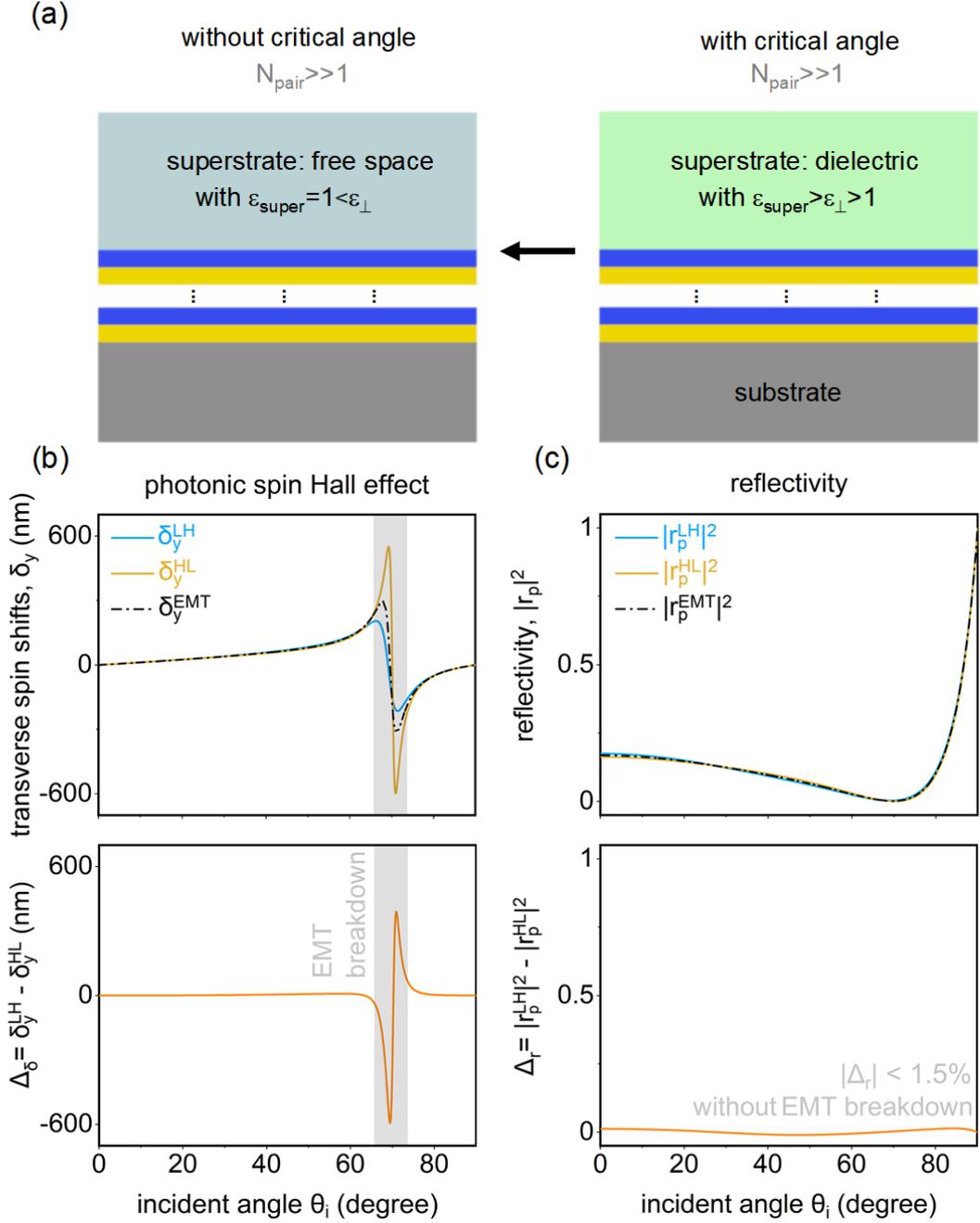

**FIG. 2. Breakdown of effective-medium theory beyond the critical angle when the superstrate is free space.** (**a**) Structural illustration. The structural setup here is the same as that in Fig. 1, except for the superstrate. In this figure, there is no critical angle, due to the absence of total reflectance. (**b**) Photonic spin Hall effect from deep-subwavelength all-dielectric multilayers. (**c**) Reflectivity from deep-subwavelength all-dielectric multilayers. The effective medium theory breakdown near the specific angle of $\theta_{i,|r_p|\to 0} = 69.5°$, when the photonic spin Hall effect is considered in (b). By contrast, the values of reflectivity from the realistic multilayers and the effective medium match well, with acceptable approximation, for arbitrary incident angles in (c).

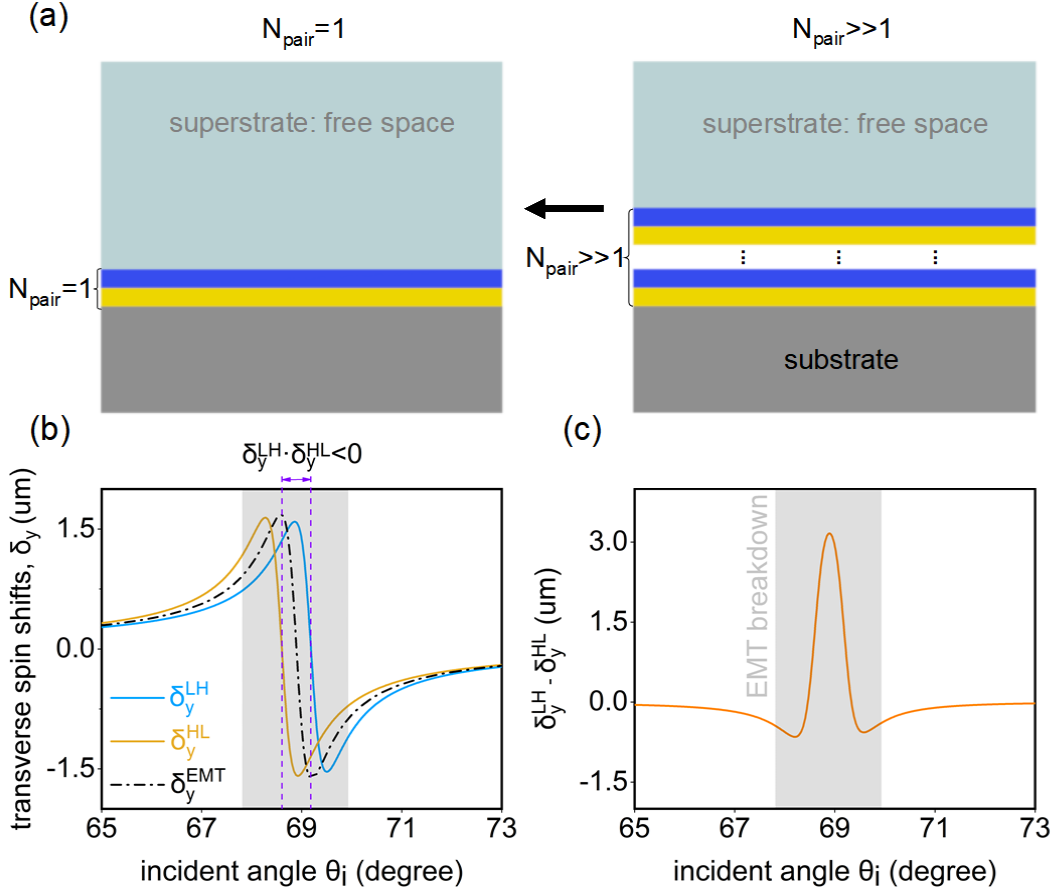

**FIG. 3. Breakdown of effective-medium theory with just two ultrathin layers.** This figure shows substantial deviations from the effective-medium theory beyond the critical angle, with a free-space superstrate, and without a periodic array of layers (namely $N_{pair}=1$). The structural setup here is the same as that in Fig. 2, except for the values of $N_{pair}$. Moreover, the substrate is chosen to fulfill $\varepsilon_{sub}=\tan^2\theta_{i,|r_p|\to 0}=6.7$, in order to maximize the sensitivity of photonic spin Hall effect on the stacking order. (**a**) Structural schematic of the deep-subwavelength all-dielectric multilayer with $N_{pair}=1$ in the left panel or $N_{pair}\gg 1$ in the right panel. (**b**) Transverse spin-shifts of photonic spin Hall effect from the realistic multilayers and the effective medium. (**c**) Deviation between the transverse spin-shifts from multilayers with different stacking orders.

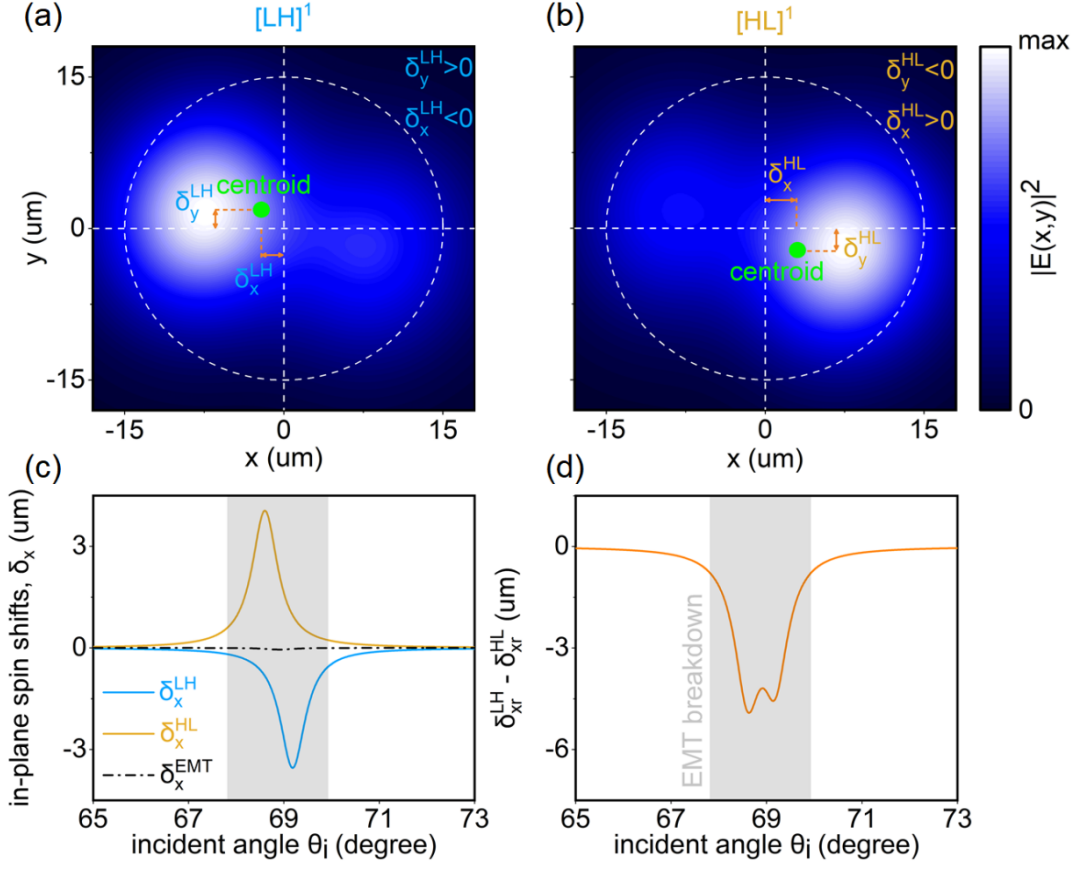

**FIG. 4. Breakdown of effective-medium theory by the in-plane spin-shifts.** The structure here is the same as that in Fig. 3 (without the critical angle, a free-space superstrate, and $N_{pair}=1$). The beam waist of the Gaussian beam is $30\lambda$. The in-plane spin-shift $\delta_x$ originated from the photonic spin Hall effect, different from the consideration of transverse shift $\delta_y$ in Figs. 1-3. (**a**, **b**) Field distributions of the reflected light with the left-handed circular polarization. Here the incident angle is $\theta_i = \theta_{i,|r_p|\to 0} = 68.9°$. (**c**) In-plane spin-shifts $\delta_x$ from the realistic multilayers and the effective medium. The directions of the in-plane spin-shift from different stacking orderings are opposite, namely $\delta_x^{HL} \cdot \delta_x^{LH} < 0$. (**d**) Deviation between the in-plane spin-shifts from multilayers with different stacking orderings.

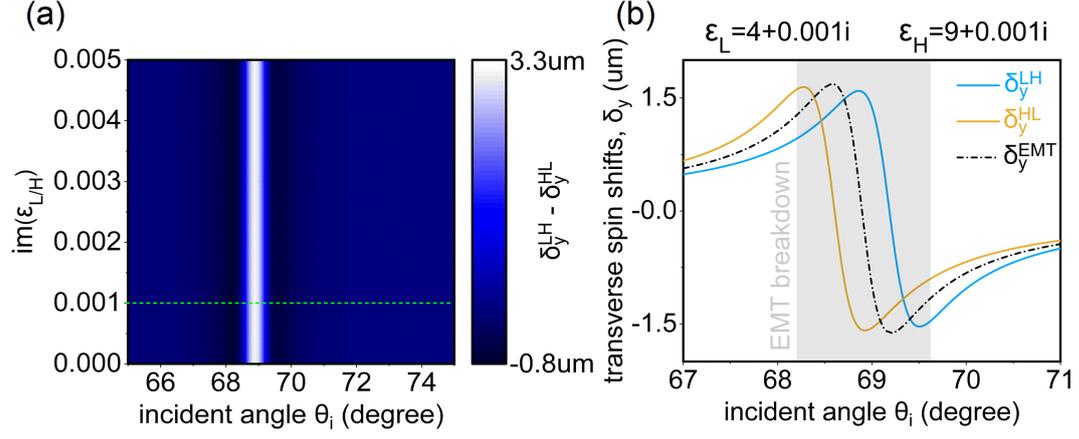

**FIG. 5. Breakdown of effective-medium theory by considering slight loss of the material.** The structure here is the same as that in Fig. 3 (without the critical angle, a free-space superstrate, and $N_{pair} = 1$), except for introducing slight loss of the dielectric layers. (**a**) Dependence of transverse spin-shift deviation from multilayers with different stacking orders on the imaginary part of dielectric constant. (**b**) Transverse spin-shifts of photonic spin Hall effect from the realistic multilayers and the effective medium with slight loss $\text{im}(\varepsilon_{L/H}) = 0.001$.